\documentclass[aps,prd,twocolumn,superscriptaddress,groupedaddress,nofootinbib]{revtex4}
\usepackage{graphicx}  
\usepackage{dcolumn}   
\usepackage{bm}        
\usepackage{amssymb}   
\usepackage{amsmath}
\usepackage{dsfont}
\usepackage{xcolor}
\usepackage{amsthm}
\usepackage{enumitem}
\usepackage{tensor}
\usepackage{diagbox} 
\usepackage{graphicx} 
\usepackage{multirow}
\usepackage{tikz}
\usetikzlibrary{shapes.geometric, arrows}
\usetikzlibrary{arrows.meta}
\usepackage{caption}
\usepackage[colorlinks]{hyperref}
\hypersetup{
     breaklinks=true,
    pdfstartview={FitH},    
    colorlinks=true,       
    linkcolor=blue,          
    citecolor=red,        
    filecolor=magenta,      
    urlcolor=blue,           
    anchorcolor=green,      
    linktocpage=true
}

\newcommand{\be}{\begin{equation}}\newcommand{\ee}{\end{equation}}
\newcommand{\bea}{\begin{eqnarray}}\newcommand{\eea}{\end{eqnarray}}
\newcommand{\brr}{\begin{array}}\newcommand{\err}{\end{array}}
\newcommand{\bit}{\begin{itemize}}\newcommand{\eit}{\end{itemize}}
\newcommand{\ben}{\begin{enumerate}}\newcommand{\een}{\end{enumerate}}

\newcommand{\ba}{\begin{array}}
\newcommand{\ea}{\end{array}}

\definecolor{darkred}{rgb}{.8,0,0}

\definecolor{darkblue}{rgb}{0,0,0.7}

\def\1{{_{1}}}\def\2{{_{2}}}

\def\noHe0{:\;\!\!\;\!\!:H_e(0):\;\!\!\;\!\!:}
\def\noHm0{:\;\!\!\;\!\!:H_\mu(0):\;\!\!\;\!\!:}
\begin{document}


\title{Generalized Eddington--Finkelstein Coordinates and Exact Vaidya-Type Solutions in Weyl Conformal Gravity}


\author{Petr Jizba}
\email[]{petr.jizba@fjfi.cvut.cz}
\author{Tereza Lehečková}
\email[]{lehecter@fjfi.cvut.cz}
\affiliation{FNSPE, Czech Technical University in Prague, Břehová 7, 115 19, Prague, Czech Republic}


\date{\today}

\begin{abstract}
We study Vaidya-type solutions in Weyl conformal gravity (WCG) using Eddington--Finkelstein-like coordinates. Our considerations focus on spherical as well as hyperbolic and planar symmetries. 
In particular, we find all vacuum dynamical solutions for the aforementioned symmetries. These are, in contrast to general relativity, structurally quite non-trivial. We provide a thorough analysis of their basic properties, such as, relation to other known WCG solutions, algebraic types, singularities, horizons, and symmetries. In the same vein, we also derive, classify, and discuss non-vacuum solutions with the Coulombic electric field and null dust. Other salient issues, such as the gauge equivalence of WCG solutions to Einstein spaces and the role of the Birkhoff--Riegert theorem, are also addressed.

\end{abstract}


\maketitle


\section{Introduction}

Einstein's General Theory of Relativity (GR) is currently the most widely accepted classical theory of gravity, demonstrating remarkable precision in predicting and explaining gravitational phenomena at scales of the solar system and smaller. However, at larger scales --- those of galaxies, galaxy clusters, superclusters, or the filamentary cosmic web --- the effectiveness of GR becomes less certain. There is strong evidence that there is a mismatch between the observed properties of the Universe and the predictions 
of general relativity when combined with the known forms of matter~\cite{Valentino:21,Abdalla:22,Skara:22}. 

One way to deal with the discrepancies within GR is to introduce an additional form of matter characterized as a nearly pressureless, perfect fluid --- dark matter. Similarly, to explain cosmological observations such as Type Ia supernovae~\cite{Riess}, the CMB~\cite{Spergel}, and the large scale structure correlations~\cite{Tangmark,Eisenstein},  which all indicate that the Universe is currently in an accelerated phase, one introduces a negative pressure energy density --- dark energy. However, it remains unclear whether dark matter and dark energy represent truly exotic matter and energy, or whether they merely reveal limitations within Einstein's theory itself. An alternative approach is to go beyond GR, e.g. to higher-derivative gravity theories, which are promising for several reasons: they circumvent Lovelock's no-go theorem (which only restricts second-order theories) and are often quantizable or at least perturbatively renormalizable~\cite{Lov:71,Lov:72,Fradkin:1985am}.

A prominent example of such a higher-derivative theory is Weyl conformal gravity (WCG), which is  a gravity theory with gauged scale invariance that has intriguing classical behavior and provides an elegant solution to the dark matter and dark energy problems~\cite{CurvE,curves}. Although its origins date back to 1918 --- shortly after the introduction of GR, with the initial goal of unifying gravity and electromagnetism~\cite{Weyl} --- it only began to attract significant attention in the 1990s.  During this period, papers began to appear (with a substantial impetus from P.D. Mannheim and his collaborators~\cite{CR-N,CurvE,curves,Mcase,mannheim_b,mannheim_c,MK2,Kazanas,mannheim,mannheim_a}) presenting various novel solutions within WCG, which gradually demonstrated that WCG has the potential not only to address the previously mentioned challenges of GR, but also to create new phenomenological predictions~\cite{mannheim,mannheim_a,JKS2,Edery:1998zi,Edery:1997hu,Edery:2001at,CP-N,PJ-GL-23}.  

In this paper we aim to study WCG solutions in the framework of Eddington--Finkelstein (E-F) type coordinates. In GR, the E-F coordinates were originally used to reformulate the Schwarzschild solution. 
They are adapted to the paths of light rays where the ingoing E-F coordinates are aligned with radially infalling light rays, while the outgoing E-F coordinates are aligned with outgoing ones. This makes the E-F coordinates particularly useful for analyzing causal structures, and thus particularly suitable for conformal gravity, where there is no mass scale present. Since the E-F coordinates are regular across the event horizon of the black hole,  they are ideal for studying local processes near the event horizon. 
In addition, E-F coordinates are much better suited for describing dynamical spacetimes, such as the Vaidya spacetime (VS), which can model, for example, a radiating black hole. This adaptability makes E-F coordinates a powerful tool for exploring time-dependent scenarios, such as black hole evaporation, accretion, or gravitational collapse~\cite{uč,EUev}.
Our goal is to explore Vaidya-type non-static solutions in WCG and investigate possible generalizations. While E-F coordinates have been used to study general properties of spherically symmetric solutions~\cite{bubbles} and the dynamics of black hole formation and evaporation in conformal gravity~\cite{Cev}, to our knowledge they have not yet been used to derive exact solutions in WCG. 

The paper is organized as follows. 
Sec.~\ref{Sec.II} introduces the E-F coordinates and the related general metric that will be used in the bulk of this paper. 
In this framework, we will also revisit the VS including some generalizations and its key features. In Sec.~\ref{Sec.III.op}, we discuss some of the fundamentals of WCG. We proceed to Sec.~\ref{Sec.IV.dc} where we find all vacuum solutions in the general E-F metric and discuss their basic properties, such as singularity and horizons, algebraic type, conformal Einstein spacetimes, conformal symmetries, and useful coordinate transformations. We also discuss the role of the Birkhoff--Riegert theorem in WCG and its relation to the solutions obtained. 
Non-vacuum solutions for Coulomb field and null dust radiation are presented in Sec.~\ref{Sec.V.bn}.  A brief summary of the results and related discussions is provided in Sect.~\ref{Sec.VI.cc}. 
For the reader's convenience, we relegate some more technical issues concerning the Weyl equivalence of Einstein spaces, as well as 
a graphical summary of the obtained WCG solutions and their main properties to two appendices.
%

\section{Eddington--Finkelstein-like coordinates and Vaidya solution \label{Sec.II}}


Eddington--Finkelstein coordinates were originally introduced to provide a more effective representation of Schwarzschild spacetime by eliminating coordinate singularities at the event horizon, allowing the study of incoming and outgoing geodesics, and providing a framework better suited for the analysis of dynamical processes such as gravitational collapse~\cite{Penrose2}.
%
Standardly, the E-F coordinates are realized by replacing the  Schwarzschild  coordinate $t$ with the retarded or advanced time, usually labeled as $v$ or $u$, respectively. The $(\theta, \varphi)$ sector, representing the unit 2-sphere, remains the same as in the Schwarzschild case.

In our following reasonings, we allow for more general E-F-like coordinates, namely coordinates with a line element of the form
\begin{equation}
\label{metrika}
H(r,w)dw^{2} \ + \ 2cdwdr \ + \ r^{2} \frac{dx^{2}+dy^{2}}{\left(1+\frac{K(x^{2}+y^{2})}{4}\right)^{2}}\, ,
\end{equation}
where $c=-1$ for retarded time and $c=1$, for advanced time (analogous notation is used e.g. in Ref.~\cite{scalarTV}). We replaced the $(\theta, \varphi) $ sector by ($ x, y $) to represent all 2-spaces with constant Gaussian curvature $ K \in
\lbrace1,0,-1\rbrace $, corresponding to spherical, Euclidean, and hyperbolic geometry, respectively. The corresponding Killing equation for the metric~(\ref{metrika}) gives 3 Killing vector fields (KVF). The general KVF has the form 
\begin{eqnarray}
\label{KV}
&&\mbox{\hspace{-10mm}}X \ = \ \ \{[K(x^{2}-y^{2})+4]c_{1} \ + \ 2Kxyc_{2} \ + \ yc_{3}
\}\partial_{x} \nonumber \\[2mm] &&\mbox{\hspace{-5mm}}+ \  \{[K(y^{2}-x^{2})+4]c_{2} \ + \ 2Kxyc_{1} \ - \ xc_{3}\}\partial_{y}\, , 
\end{eqnarray}
where $c_{i}$ are constants. 

E-F coordinates are often useful for describing dynamical spacetimes, which can otherwise be quite complicated to represent in Schwarzschild coordinates~\cite{uč}. Here we will focus in particular on VS. In GR, the latter represents a dynamical version of the Schwarzschild spacetime, characterized by a dynamical mass and a pure (incoming or outgoing) radiation energy-momentum tensor. It is used to model radiating stars, semiclassical black hole evaporation, radiating white holes, spherically symmetric gravitational collapse, etc.~\cite{uč, Vhorizons, EUev, Vgeom}. There is also a charged version~\cite{Venergy} (sometimes referred to as Vaidya--Bonnor spacetime) and a version with the cosmological constant $\Lambda $, which can both be written together in a compact form by using the metric~(\ref{metrika}) with
\begin{eqnarray}
&&K \ = \ 1\, , \nonumber \\[2mm]
&&H(r,w) \ = \ -\left(1-\frac{2m(w)}{r}+\frac{q^{2}(w)}{r^{2}}-\frac{\Lambda}{3} r^{2}\right). \ ~~~~~
\label{V}
\end{eqnarray}
For the sake of brevity, we will call this VS in the following text, although it is actually an extension of it by charge and a cosmological constant. This spacetime contains a Coulomb field and pure radiation of the null-dust type. In particular, the latter generally represents a field (or particles with zero rest mass) propagating at the speed of light and with an energy-momentum tensor with a non-zero component 
\begin{eqnarray}
T^{M}_{w w} \ = \  c[ m'(w)r-q(w)q'(w)]/r^{3}\, , 
\end{eqnarray}
see e.g.~\cite{uč}. 
The functions $ m(w) $ and $ q(w) $ can be interpreted as mass and electric charge, respectively. The case with even more general  $m = m(r,w)$ is sometimes called the generalized Vaidya solution~\cite{genV}.

The Vaidya spacetime is of Petrov D type, has a singularity at $r=0$ and a trapping or apparent (for the asymptotically flat case) horizon at $H(r,w)=0$.  To meet the energy conditions, the radiation must fulfill the inequality $T^{M}_{w w} \geq 0$. 
%
%
This does not seem to be satisfied in some cases, but this fact is caused by neglecting the Lorenz force~ rather than being a real problem~\cite{Venergy}. We will encounter a similar situation in Sec.~V, since our choice of the stress-energy tensor there will be inspired by VS in conventional GR.
The analysis of event horizons is generally quite complicated in dynamical spacetimes~\cite{Cev, CKVt}, and so is the ensuing thermodynamics, though it may be simplified in some cases. 

If we specifically assume the linear mass function  $m(w)=a( A_{1}w+A_{2})$, the linear charge function $q(w)=b(A_{1}w+A_{2})$  and $\Lambda=0$, the resulting VS~(\ref{V}) admits a {\em conformal} Killing vector field (CKVF)   
\begin{equation}
\label{KVF}
X^{\rm c} \ = \ (A_{1}w+A_{2})\partial_{w} \ + \ A_{1} r\partial_{r}\, .
\end{equation}
Here we recall that CKVFs are a generalization of KVFs which preserve the metric tensor up to a scalar factor --- CKVF are thus generators of {\em conformal isometry}.
The linear mass function is also one of the cases where transformation to double-null coordinates is possible~\cite{DoubleN} and is used as a toy model for collapse and evaporation~\cite{Cev}.
By Weyl rescaling, it is possible to get a static spacetime which has CKVF~(\ref{KVF}) as KVF. The latter allows one to study thermodynamic and causal properties in this new spacetime and then convert the results obtained to the original VS~\cite{Vhorizons, CKVt}. For instance, the event horizon in static spacetime is collocated with one of the Killing horizons of (\ref{KVF}). Since the causal properties are conformally invariant, this location can be transformed back to the original spacetime. We will see this in more detail in Sec.~\ref{Sec.IV.dc}. 

Some authors have briefly considered the possibility of different $K$, see e.g. Ref.~\cite{scalarTV}, which could potentially represent a dynamical counterpart of so-called topological black holes, i.e. black holes whose event horizons are surfaces of non-trivial topology~\cite{uč, top}. To our knowledge, there has been no deeper mathematical or physical analysis of these solutions.

\section{Weyl conformal gravity \label{Sec.III.op}}

For the sake of logical consistency, we will give here a brief overview of the most important aspects of  WCG that are relevant to our subsequent analysis. Further details can be found, e.g., in Refs.~\cite{RJ,Bach}.

WCG is a pure metric theory invariant under diffeomorphisms as well as Weyl rescaling of the metric tensor by smooth, non-singular real functions $ \Omega(x)$ without zero points: $g_{\mu\nu}(x) \rightarrow \Omega^2(x)g_{\mu\nu}(x)$. The simplest WCG action in four-dimensional spacetime can be written in the form~\cite{Bach}
\begin{eqnarray}
S_{\text{w}}[g] \ = \ -\frac{1}{4G_{\text{w}}^2} \int d^4x \sqrt{|g|} C_{\mu\nu\rho\sigma} C^{\mu\nu\rho\sigma}\, .
\label{III.4.cv}
\end{eqnarray}
Here  $C_{\mu\nu\rho\sigma}$ is the
{Weyl tensor} which can be rewritten as
\begin{eqnarray}
C_{\mu\nu\rho\sigma}  \ \!&=&\ \!
R_{\mu\nu\rho\sigma} \ - \ \left(g_{\mu[\rho}R_{\sigma]\nu} \ - \ 
g_{\nu[\rho}R_{\sigma]\mu} \right)
\nonumber\\[1mm]
&+& \ \! \frac{1}{3}R \ \! g_{\mu[\rho}g_{\sigma]\nu}\, ,
\label{P2a}
\end{eqnarray}
with $R_{\mu\nu\rho\sigma}$ being the {Riemann curvature tensor},
$R_{\mu\rho}=R_{\mu\nu\rho}{}^\nu$  the  {Ricci tensor}, and $R= R_\mu{}^\mu$
the {scalar curvature}.  The Weyl tensor has all the algebraic properties
of the Riemann curvature tensor, i.e. 
\begin{eqnarray}
&&C_{\alpha \beta \gamma \delta} \ = \  C_{[\alpha \beta][\gamma \delta]} \ = \  C_{\gamma \delta \alpha \beta}\, ,\nonumber \\[2mm] 
&&C_{\alpha[\beta\gamma\delta]} \ = \ 0\, ,
\end{eqnarray}
and in addition is trace-free, i.e. $C_{\alpha \beta \gamma}^{~~~~\alpha} = 0$. 
Weyl tensor is identically zero in three dimensions.
The action~(\ref{III.4.cv}) is Weyl invariant in four dimensions and mimics a Yang--Mills structure with the dimensionless coupling constant $G_{\text{w}}$. This makes it power-counting renormalizable.

Throughout this work, we adopt the spacelike metric signature $(-,+,+,+)$. The sign convention for the Riemann curvature tensor is chosen so that
\begin{eqnarray}
R^{\alpha}_{~\beta \mu \nu} \ \!&=& \ \! \partial_{\mu} \Gamma^{\alpha}_{~\beta \nu} \ - \ \partial_{\nu} \Gamma^{\alpha}_{~\beta \mu} \nonumber \\[2mm] &+& \ \!\Gamma^{\alpha}_{~\kappa \mu} \Gamma^{\kappa}_{~\beta \nu} \ - \ \Gamma^{\alpha}_{~\kappa \nu} \Gamma^{\kappa}_{~\beta \mu} \, .~~~~
\end{eqnarray}
For notational simplicity, we use the following conventions: \( R^2_{\mu\nu\rho\sigma} \) represents the square of the Riemann tensor, \( R_{\mu\nu}^2 \) denotes the square of the Ricci tensor, and \( R^2 \) is the square of the scalar curvature. The square of the Weyl tensor, denoted by \( C^2 \), expands in four dimensions as
\begin{eqnarray}
C^2 \ = \  R_{\mu\nu\rho\sigma}^2 \ - \  2R_{\mu\nu}^2 \ + \  \frac{1}{3}R^2\, .
\end{eqnarray}
An important related invariant is the Gauss--Bonnet (GB) term \( E_4 \), given as
\begin{eqnarray}
E_4 \ = \  R_{\mu\nu\rho\sigma}^2 \ - \  4R_{\mu\nu}^2 \ + \  R^2\, ,
\end{eqnarray}
which serves as the integrand for the topological Euler--Poincaré invariant.
Using the Chern--Gauss--Bonnet theorem, we can recast the Weyl action equivalently (up to a topological term) as
\begin{eqnarray}
\label{akce}
S^{\text{w}}[g] \ = \  -\frac{1}{2G_{\text{w}}^2} \int d^4x \sqrt{|g|} \left(R_{\mu\nu}^2 - \frac{1}{3} R^2\right).
\end{eqnarray}
The variation of $S^{\text{w}}$ with respect to the metric yields the dynamical equation, the so-called Bach vacuum equation
\begin{eqnarray}
\label{vacuum}
B_{\mu\nu} \ \equiv  \ \nabla^\kappa \nabla^\lambda C_{\mu\kappa\nu\lambda} - \frac{1}{2} R^{\kappa\lambda} C_{\mu\kappa\nu\lambda} \ =  \ 0\, ,
\end{eqnarray}
where $B_{\mu\nu}$, the {\em Bach tensor}, is symmetric and trace-free. 
In addition, since 
\begin{eqnarray}
B^{\mu\nu} \ \propto \ \frac{1}{\sqrt{-g}} \ \! \frac{\delta S^{\text{w}}}{\delta g_{\mu\nu}}\, ,
\end{eqnarray}
we see that under the Weyl transformations of the metric
\begin{eqnarray}
B^{\mu\nu} \ \mapsto \ \Omega^{-6} B^{\mu\nu}\, ,
\end{eqnarray}
which often allows to dramatically simplify the solution of the Bach equation. Solutions to  Eq.~(\ref{vacuum}) are known as \emph{Bach-flat} backgrounds. Notably, every Einstein space (i.e. spacetime where \( R_{\alpha \beta} = \Lambda g_{\alpha \beta} \)) satisfies Bach's equation, see e.g.,~\cite{CP-N}, although the converse is not true. We will have more to say on this issue in Sec.~\ref{IV.B.bb}.

In the presence of other fields (non-vacuum case), the Bach equation generalizes to
\begin{equation}
\label{nvac}
B_{\mu \nu} \ = \ \frac{G_{{\text{w}}}^{2}}{2} T_{\mu \nu} \, .
\end{equation}
The latter implies that the stress-energy tensor in WCG must be traceless, which in turn requires the matter-field action to be Weyl invariant, so the tracelessness condition immediately excludes kinematical or mechanical masses. Strictly unbroken Weyl symmetry therefore does not allow for mass scales. However, if the scale symmetry is spontaneously broken (when quantized), then it is possible to generate mass scales, and  this can be done while maintaining the tracelessness condition on the matter energy-momentum tensor~\cite{MK94}.

As already mentioned, all  Einstein spaces are also solutions of the Bach vacuum equation, while the opposite is not true.  In fact, there may be additional terms or even major changes in the Bach vacuum solutions~\cite{CR-N,CurvE}. A paradigmatic example is the M-K solution~\cite{CR-N}, which is the WCG analog of the exterior Schwarzschild solution in GR, i.e. the most general spherically symmetric static Bach  vacuum solution. In the general  E-F-like coordinates~(\ref{metrika}) it reads
\begin{eqnarray}
\label{M-K}
&&\mbox{\hspace{-8mm}}K  \ = \  1\, , \nonumber \\[2mm] 
&&\mbox{\hspace{-8mm}}H(r)  \ = \  -\left[1 -  3\beta \gamma  -  \frac{\beta(2-3\beta\gamma)}{r} \ + \ \gamma r  -  \kappa r^{2} \right] ,~~
\end{eqnarray}
%
%
where $ \beta, \: \gamma, \: \kappa $ are constants. Here $\beta$ represents the mass of the source, $\gamma$ is related to the dark matter, and $\kappa$ has an effect of the cosmological constant. Versions with $K=-1 $ and $K=0 $, i.e. the WCG analogues of topological black holes, have also been found, see e.g. Ref.~\cite{Ctop}. 

One might naturally ask whether these additional (non-Einstein space) solutions might not be at least Weyl related to some Einstein spaces. And if so, how does the situation generalize to non-vacuum Einstein spacetimes? While there are topologically non-trivial Bach vacuum solutions that cannot be mapped to Einstein spaces by a Weyl transformation~\cite{CP-N}, many solutions allow such a mapping. In particular, for the M-K solution, the linear term in~(\ref{M-K}) can be eliminated by an appropriate Weyl rescaling~\cite{Schmidt}, yielding a spacetime equivalent to Schwarzschild--(anti)-de Sitter spacetime. However, the required conformal factor is singular at certain points. As a result, the equivalence between these spacetimes is only local, and some global spacetime properties, such as the Noether--Wald charge or the homotopy groups, are often different~\cite{N-Wmass}.  Similarly, a Birkhoff--Riegert theorem for WCG, states that all spherically symmetric electrovacuum solutions in WCG can be Weyl transformed into the (charged) M-K solution~\cite{Riegert}. Again, such transformations often involve singularities, so the equivalence is only local. We will say more about this in Sec.~\ref{IV.C.cc}.

As an aside, we note that there is an interesting difference between the charged version of (\ref{M-K}) and its GR counterpart (Reissner--N\"{o}rdstrom solution). If we add the Coulomb field $ F_{wr}=q/r^{2} $ to the Schwarzschild solution, the extra term that appears in the metric is of order $ 1/r^{2} $. If we do the same in WCG, the extra term is of order $ 1/r $, resulting in a different structure of spacetime~\cite{CR-N,CP-N}.

\section{Vacuum case \label{Sec.IV.dc}}

Now we will solve and analyze the WCG field equations for the ansatz metric~(\ref{metrika}). We will start with the vacuum case, i.e. the Bach equation~(\ref{vacuum}). In GR there are no dynamical vacuum solutions for~(\ref{metrika}). As can be seen from the Einstein tensor components~(\ref{Einstein}) in the Appendix A, the most general GR vacuum solution is Schwarzschild--(anti)-de Sitter, which is static. In this section, we will see that the situation in WCG is quite different. 

Here and throughout, we will use the following notational convention: if a function $F(r,w)$ depends on both $w$ and $r$, we denote the derivative with respect to $r$ by $F'$ and the derivative with respect to $w$ by $\dot{F}$. If the function $F(r,w)$ depends on only one variable, i.e., $F(r,w) \equiv F(r)$ or $F(r,w) \equiv F(w)$, we denote its derivative by $F'$, regardless of which variable it is related to.

\subsection{General solution }

Although the Bach tensor in the metric~(\ref{metrika}) is quite complicated, it can be simplified considerably by solving~(\ref{vacuum}) for the $rr$-component first. In this case, we get the Bach equation  
\begin{eqnarray}
\label{Brr}
&&\mbox{\hspace{-12mm}}B_{rr}\ = \ -\frac{1}{6r} \left[r \ \!\frac{\partial^{4} H(r,w)}{\partial r^{4}} \  + \ 4 \ \! \frac{\partial^{3} H(r,w)}{\partial r^{3}}\right] \ = \ 0\, ,
\end{eqnarray}
which has a generic solution
\begin{equation}
\label{H}
H(r,w) \ = \ F_{1}(w) \ + \ \frac{F_{2}(w)}{r} \ + \ F_{3}(w)r \ + \ F_{4}(w)r^{2}\, .
\end{equation}
Substituting this back into the remaining components of the Bach tensor, we find that the only non-zero components are
%
\begin{widetext}
\begin{eqnarray}
-6cr^{4}B_{wr} &=& K^{2} \ + \ 3F_{2}F_{3} \ - \ F_{1}^{2} \ + \ 6cF_{2}'\, , \quad B_{xx} \ = \ B_{yy} \ = \ -\frac{16cr^{2}B_{rw}}{[K(x^{2}+y^{2})+4]^{2}}\, ,\ \label{Bwr} \\[2mm]
-6r^{5}B_{ww}&=&F_{2}[K^{2} \ + \ 3F_{2}F_{3} \ - \ F_{1}^{2} \ + \ 6cF_{2}'] \ + \ rF_{1}[K^{2} \ + \ 3F_{2}F_{3} \ - \ F_{1}^{2} \ + \ 6cF_{2}']
\nonumber \\[2mm]
&&+ \ r^{2}[K^{2}F_{3} \ - \ F_{3}F_{1}^{2} \ + \ 3F_{2}F_{3}^{2} \ - \ 2cF_{1}'F_{1} \ + \ 3cF_{3}'F_{2} \ + \ 9cF_{3}F_{2}' \ + \ 6F_{2}'']
\nonumber \\[2mm]
&&+ \ r^{3}[K^{2}F_{4}\ - \ F_{4}F_{1}^{2} \ + \ 3F_{2}F_{3}F_{4} \ + \ 3cF_{2}F_{4}'\ - \ cF_{3}F_{1}'\ + \ 12cF_{4}F_{2}'\ + \ 2F_{1}'']\, .
\label{Bww}
\end{eqnarray}\\
\end{widetext}
We have omitted explicitly writing multiplication by $c^2$ terms, since $c = \pm 1$. This convention, along with the fact that $1/c=c$, will be used in the following formulas.
As~(\ref{Bww}) is a 3rd order polynomial in $r$, it leads to four equations. A fifth equation follows from~(\ref{Bwr}). 

Let us now solve these equations. There will be two branches of solutions. {\em The first branch} corresponds to $F_{2} (w)=0$. In this case, $B_{rw}=0$ implies that $F_{1}(w)=\pm K$. After inserting this into $B_{ww}$, the latter will be identically zero, so the first solution branch is
\begin{equation}
\label{vacback}
 F_{1}(w)\ = \ \pm K, \;\; 
 \; F_{2}(w)\ = \ 0\, , 
\end{equation}
and $F_{3}(w)$ and $F_4(w)$ are arbitrary. 
{\em The second branch} corresponds to $F_{2}(w) \neq 0$. In this case, we can express $F_{3}(w)$ from $ B_{rw}=0 $. This will automatically set the 0th, 1st and 2nd order in $ r $ in (\ref{Bww}) to zero. So we are left with the last equation (from the 3rd order in $ r $). This can be understood as a first-order linear differential equation for $ F_{4}(w) $. We can solve it, yielding the second-branch solution
\begin{widetext}
\begin{eqnarray}
\label{vacfull}   
&&F_{3}(w) \ = \ \frac{-K^{2} \ + \ F_{1}^{2}(w) \ - \ 6cF_{2}'(w)}{3F_{2}(w)}\, , \quad
F_{4}(w) \ = \ \frac{-\frac{K^{2}F_{1}(w)}{9} \ + \ \frac{F_{1}^{3}(w)}{27} \ - \ \frac{2cF_{1}'(w)F_{2}(w)}{3} \ + \ C_{1}}{F_{2}^{2}(w)}\, ,
\end{eqnarray}\\
\end{widetext}
where $F_{1}(w)$ and $ F_{2}(w) \neq 0$ are arbitrary and $ C_{1} $ is an integration constant. It is quite remarkable that the function $F_4$ is completely explained by only two functions and one constant without any integration involved. Equations~(\ref{vacback})-(\ref{vacfull}) are the main new results of this section. The case where $F_{2}(w)=0$ only at some discrete points $w$ formally behaves like~(\ref{vacfull}), but diverges at those points. 

\subsection{Basic properties of the general solution \label{IV.B.bb}}

Now, we will analyze some basic properties of the previous solutions to get some basic understanding of their structure and meaning. We start with the relationship between~(\ref{vacfull}) and the M-K solution~(\ref{M-K}). It can be seen that if we fix our $F_{i}(w)$ functions in (\ref{vacfull}) to be constants and set $K=1$, we get the M-K solution~(\ref{M-K}) in E-F-like coordinates. In our notation we have $\kappa =F_{4}$, {$\gamma= - F_{3}$} and {$\beta=- (1+F_{1})/3F_{3}$}. As a consistency check one can easily verify that with this identification, $F_2(w) =  \beta (2-3\beta\gamma)$ (that is needed for the M-K solution) automatically satisfies~(\ref{vacfull}). Thus, we see that the M-K solution belongs to an infinite family of solutions found here, parametrized by two functions $F_{1}(w)$ and $F_{2}(w)$ and a constant $C_1$. Analogous reasoning also holds for topological black holes~\cite{Ctop}, which have a similar structure of $H(r)$ as the M-K solution, just with renamed constants. 
Thus, for any given functions $F_1$ and $F_2$, the solution (\ref{vacfull}) could generally describe a (possibly topological) M-K type dynamical black or white hole, or a naked singularity (or even something more exotic for the negative mass term).

As for the solution~(\ref{vacback}), this is parameterized by two functions, $F_3(w)$ and $F_4(w)$. In the case where $F_3(w) = 0$ and $F_4(w)$ is a non-zero constant, we have a de Sitter solution, which in conventional GR requires the presence of a cosmological constant.  If, in addition, $F_3(w)$ is also a non-zero constant, such a constant has the dimension of the acceleration, and thus the solution of the Bach vacuum equation dynamically generates a characteristic constant acceleration without having introduced one in the Lagrangian. For non-constant $F_3(w)$ and $F_4(w)$, the solution~(\ref{vacback}) can be interpreted as a dynamical Mach-type background, since there is no ``mass'' term corresponding to $F_{2}(w)$. By the ``mass'' term we mean the local gravitational source term in the sense of the M-K or topological black hole solution.

To get a deeper grip on the solution, we choose the null tetrad $\lbrace n^{\mu}, l^{\mu}, m^{\mu}, \bar{m}^{\mu} \rbrace$, where the first two members are real and the last two are complex and conjugate to each other. All tetrads are perpendicular to each other except $l^{\mu}n_{\mu}=-1$ and $m^{\mu} \bar{m}_{\mu}=1$. These null tetrads are also used as the basis of the Newman--Penrose formalism, and in WCG they were recently employed in Ref.~\cite{CP-N}. For the line element~(\ref{metrika}) it is convenient to choose
\begin{eqnarray}
\label{tetrad}
&&m \ = \  \frac{K(x^{2}+y^{2})+4}{4\sqrt{2}r} \partial_{x} \ + \  \frac{i[K(x^{2}+y^{2})+4]}{4\sqrt{2}r}\partial_{y\,} , \nonumber \\[2mm]
&&l \ = \ \partial_{w}-\frac{cH(r,w)}{2}\partial_{r}\, , \quad n \ = \  -c\partial_{r} \, .
\end{eqnarray}
The curvature is encoded in the Newman--Penrose formalism in five complex Weyl scalars~\cite{Chandra}
\begin{eqnarray}
    &&\mbox{\hspace{-4mm}}\Psi_0 \ = \ l^{\mu}m^{\nu}l^{\alpha}m^{\beta}\tensor{C}{_\mu_\nu_\alpha_\beta}\, , \;\;\;\;  \Psi_1 \ = \ l^{\mu}n^{\nu}l^{\alpha}m^{\beta}\tensor{C}{_\mu_\nu_\alpha_\beta}\, , \nonumber \\[2mm]
&&\mbox{\hspace{-4mm}}\Psi_2 \ = \ l^{\mu}m^{\nu}\bar{m}^{\alpha}n^{\beta}\tensor{C}{_\mu_\nu_\alpha_\beta}
    \, , \;\;\;\;
    \Psi_3 \ = \ l^{\mu}n^{\nu}\bar{m}^{\alpha}n^{\beta}\tensor{C}{_\mu_\nu_\alpha_\beta}\, ,\nonumber \\[2mm]
&&\mbox{\hspace{-4mm}}\Psi_4 \ = \ n^{\mu}\bar{m}^{\nu}n^{\alpha}\bar{m}^{\beta}\tensor{C}{_\mu_\nu_\alpha_\beta}\, .
\label{III.16.cc}
\end{eqnarray}
It can be checked that by using~(\ref{tetrad}) and the metric (\ref{metrika}) with (\ref{H}), the only non-zero Weyl scalar is 
\begin{equation}
\Psi_{2} \ = \ -\frac{[K \ + \ F_{1}(w)]r \ + \ 3F_{2}(w)}{6r^{3}}\, .
\label{24.cc}
\end{equation} 
The latter implies that all the ensuing solutions are of Petrov type~D, if $\Psi_{2} \neq 0$, or Petrov type~O (i.e. conformally flat), if $\Psi_{2} = 0$, see e.g.~Ref.~\cite{Chandra} for details of the Petrov classification. The above Weyl scalars transform trivially as $\Psi_i\rightarrow\Omega^{-2}\Psi_i$. This shows that the Petrov types (which are formulated in terms of Weyl scalars) are preserved by Weyl transformations, and that the restrictions of the Bach equations on individual Petrov types are also Weyl invariant.

From~(\ref{24.cc}), we also see that $r=0$ is indeed the genuine singularity since, for example, one of the complex scalar polynomial invariants~\cite{uč}
\begin{eqnarray}
\label{I}
I \ = \ \Psi_{0} \Psi_{4} \ - \ 4 \Psi_{1}\Psi_{3} \ + \ 3\Psi_{2}^{2} \ = \  3\Psi_{2}^{2}\, ,
\end{eqnarray}
diverges for $F_{2}(w) \neq 0$, $F_{1}(w) \neq -K$. So, the singularity is preserved under diffeomorphisms. In addition, we can use the fact that the Weyl squared term is not only a scalar density, but in four dimensions it is also invariant under Weyl transformations. In particular, if $C^2$ is singular in one spacetime, then there is no non-singular conformal factor $\Omega$, which would remove the singularity in another  Weyl-equivalent spacetime. In our case, we have
\begin{eqnarray}
\label{C2}
C^2 \ = \ \frac{[2r(K \ + \ F_1(w)) \ + \ 6F_2(w)]^2}{3r^6}\, ,
\end{eqnarray}
which is singular only at the point where $r = 0$ (provided $F_{1}(w) \neq -K$, $ F_{2}(w) \neq 0$ and functions $F_{1}(w)$ and $F_{2}(w)$ are nonsingular). 
Note that in the above invariants there is a singularity at $r=0$ even for the case $F_2(w)=0$, providing $F_1(w) \neq -K$, which is a case where there is no coordinate singularity in (\ref{metrika}). In contrast, the divergent points in the solution (\ref{vacfull}) for the case where $F_2(w)=0$ in discrete points seem to be only coordinate singularities, since these points are not problematic in the invariants.

Although event horizon localization can be pretty complicated in dynamical spacetimes~\cite{Vhorizons,apparent,genV}, we can find trapping horizons quite easily. In particular,  one can employ ingoing and outgoing null geodesic expansions to find and classify various horizon surfaces~\cite{apparent}. By the latter we mean expansions of $n^{\mu}$ and  $l^{\mu}$ from  (\ref{tetrad}), computed as $\Theta_{n}=h^{\mu \nu}n_{\nu; \mu}$, $ \Theta_{l}=h^{\mu \nu}l_{\nu; \mu}$, where $ h^{\mu \nu} =g^{\mu \nu}+n^{\mu}l^{\nu}+l^{\mu}n^{\nu}$. For~(\ref{metrika}) these are
\begin{equation}
\Theta_{n} \ = \ -\frac{2c}{r}\, , \quad \Theta_{l} \ = \ -\frac{cH(r,w)}{r}\, .
\end{equation}
Using the notation of Ref.~\cite{horizons}, the trapping horizon(s) are defined by three conditions, namely
\begin{eqnarray}
\Theta_{l}|_{TH} \ = \ 0\, , \;\;\; \Theta_{n}|_{TH} \ \neq \ 0\, , \;\;\; \mathcal{L}_{n} \Theta_{l}|_{TH} \ \neq \ 0\, ,  
\label{28.jk}
\end{eqnarray}
where $\mathcal{L}_{n}$ is the Lie derivative along a vector $n$.
Note that the first two conditions in~(\ref{28.jk}) are satisfied for $H(r,w)=0 $, which is also consistent with the placement of the horizons in the standard VS in GR~(\ref{V}). The last condition can be explicitly written as
\begin{eqnarray}
n^{\mu} \partial_{\mu} \Theta_{l}\mid_{TH} \ = \ (rH'-H)/r^{2}|_{TH} \ \neq \ 0\, ,
\end{eqnarray}
which means that $H(r,w)=0$ is a trapping horizon as long as $H'(r,w) \neq 0$ at the same time. 
One can further distinguish between a past trapped horizon, for which $\Theta_{n}|_{TH} >0$, and a future trapped horizon for which $\Theta_{n}|_{TH} <0$. Similarly one can differentiate between and outer horizon if $\mathcal{L}_{n} \Theta_{l}|_{TH} <0$,  and inner horizon if $\mathcal{L}_{n} \Theta_{l} |_{TH} >0$. In the context of black holes, the most pertinent case is that of the future outer trapping horizon. In this case, the preceding classification encapsulates the idea that the ingoing light rays should be converging,  $\Theta_n|_{TH} <0 $, and the outgoing light rays should be instantaneously parallel on the horizon, $\Theta_l|_{TH} =0$, and diverging just outside the horizon and converging just inside, $\mathcal{L}_{n} \Theta_{l}|_{TH}  \neq  0$. In the context of white holes, past outer horizons also appear. In addition, future and past inner horizons appear in contracting and expanding cosmologies, respectively.

In our case (\ref{H}), the equation $H(r,w)=0$ can have (maximally) three $r$-roots, which are generically $w$-dependent (though not necessarily all physical, similarly as in the M-K case~\cite{OnM-K}), corresponding to black/white hole and cosmological horizons. 
Thus, our rather general analysis reveals that in WCG, at least some of the Bach vacuum spacetimes change horizon position in time, and do so without radiating any matter field, which is impossible in GR. In particular, note that the Weyl scalar $\Psi_4 = 0$, which is tantamount to the zero radiation field. This may seem to contradict the Birkhoff--Riegert theorem, which states that all spherically symmetric electrovacuum solutions in WCG are Weyl equivalent to the (charged) M-K solution~\cite{Riegert}.  However, the theorem has an important caveat, namely that it is not easily applicable to singular Weyl transformations, so that in many cases the equivalence is at best local. In fact, singular Weyl transformations typically lead to spacetimes that are globally Weyl inequivalent, i.e., they differ in various global characteristics, such as the Noether--Wald charge, homotopy groups, Betti numbers, etc.~\cite{CP-N,N-Wmass}.    
The point is that such transformations in general change the identification in the
covering space. 
%
%
We shall comment on this point more in the following subsection.

 \subsection{Birkhoff--Riegert theorem  and Weyl transformations \label{IV.C.cc}}

The Weyl inequivalence between spacetimes is crucial for the classification of solutions in WCG. In fact, since in WCG the scale transformation is gauged,  WCG is a theory of equivalence classes, where each class consists of the Weyl equivalent spacetimes.  While the causal properties
of all spacetimes in a given class are the same, different classes have different causal structures.

In this subsection we will study spacetimes that are Weyl equivalent to our solutions encapsulated by Eqs.~(\ref{vacback})-(\ref{vacfull}). To this end, we begin by commenting on how these spacetimes relate to Riegert's result concerning the uniqueness of the M-K solution. 
In~\cite{Riegert}, Riegert first considered the most general spherically symmetric metric 
\begin{eqnarray}
ds^{2} &=& g_{ij}(x^{0}, x^{1})dx^{i}dx^{j} \nonumber \\[2mm] 
&&+ \ l(x^{0}, x^{1})[d \theta^{2} \ + \ \sin(\theta)d\varphi^{2}]\, , 
\end{eqnarray}
with $i,j  =   0,1$.
Using the Weyl transformation with $\Omega^{2}(x^{0}, x^{1})^{2}=l^{-1}(x^{0}, x^{1})$, he then showed that it is possible to obtain a metric with additional non-null KVF, which can then be used to coordinate transform the obtained metric into an explicitly static one.  We now consider the metric~(\ref{metrika}), which in a compact form describes both spherically, planarly and hyperbolically symmetric spaces in the E-F-like coordinates. When using~(\ref{H}) together with~(\ref{vacfull}), the metric (\ref{metrika}) has a general CKVF of the form
\begin{equation}
\label{CKVFg}
    \tilde{X}^{\rm c} \ = \ 6F_{2}(w)\partial_{w} \ + \ 2r[rF_{1}'(w) \ + \ 3F'_{2}(w)] \partial_{r}\, .
\end{equation} 
Note that CKVF also exists for~(\ref{vacback}), but it is trivial. Dividing (\ref{metrika}) by $r^{2}$, the above CKVF becomes KVF for the transformed metric
\begin{equation}
\label{metrikaCT}
\frac{H(r,w)}{r^{2}}dw^{2} \ + \ \frac{2cdwdr}{r^{2}}  \ +  \frac{dx^{2}+dy^{2}}{\left(1+\frac{K(x^{2}+y^{2})}{4}\right)^{2}}\, ,
\end{equation}
which now consists of two independent 2-spaces.  It can be checked that (\ref{CKVFg}) is exactly the KVF from Riegert's proof, which can be written in the form 
\begin{eqnarray}
\tilde{X}^{\rm c}_i \ = \ \epsilon^{\ \!j}_i \partial_j \tilde{R} \, , 
\label{35.vbn}
\end{eqnarray}
where $i,j=0,1$, $x^{0},x^{1} \equiv w,r$, $\epsilon_{ij}$ is an antisymmetric Levi-Civita tensor and $\tilde{R}$ is the Ricci scalar of $(w,r)$ 2-space.  In fact, for (\ref{metrikaCT}) with the general solution~(\ref{H}) we have that
\begin{eqnarray}
\tilde{R} \ = \ 2[r F_1(w) \ + \ 3F_2(w)]/r \, , 
\end{eqnarray}
so (\ref{CKVFg}) matches the form~(\ref{35.vbn}), independently on $K$.   From this we can see that $\tilde{R}$ is constant for (\ref{vacback}), which is also consistent with Riegert's procedure --- that is, for (\ref{vacback}), the $(w,r)$ 2-space is a maximally symmetric space of constant curvature. Such a space admits three KVFs, at least one of which is non-null. So for both branches~(\ref{vacback}) and (\ref{vacfull}) we have at least one non-null KVF. In such a case, a coordinate transformation can be made so that the metric~(\ref{metrikaCT}) becomes explicitly static (non-null KVF explicitly indicates time-independence)~\cite{Riegert}. Since the most general static solution is M-K (\ref{M-K}) or WCG topological black hole (depending on $K$), solutions are conformally equivalent to it. However, it should be emphasized again that the proven equivalence is true only locally, as the Weyl transformation $\Omega^{2}=1/r^{2}$ is singular.

In passing, we can mention that the conformal transformation from Riegert's procedure is not necessarily the only possible one. For example, for~(\ref{CKVFg}) any Weyl transformation of the form
\begin{eqnarray}
    \label{OmT}
    \Omega(w,r) \ = \ \frac{1}{r} \ \!G\!\left(\frac{rF_1(w)+3F_2(w)}{r} \right) ,
\end{eqnarray}
will make it KVF. This transformation may be regular for some $F_i(w)$ cases and suitable choices of function $G$, but there is a large class of solutions for which it is not. 

To give an example of the outlined procedure, but also to see an interesting relation between WCG and GR solutions, let us discuss the ``linear mass case'' (LMC). By analogy with the situation in GR, we can ask whether there are circumstances under which~(\ref{KVF}) can represent CKVF for solutions~(\ref{vacback}) and (\ref{vacfull}). Although (\ref{KVF}) can clearly be a special case of (\ref{CKVFg}), the circumstances under which it is a CKVF are broader. To see this, we take (\ref{KVF}) as an ansatz, and for the general metric~(\ref{metrika}) we obtain
\begin{eqnarray}
X^{\rm c}_{\mu;\nu} \ +X^{\rm c}_{\nu;\mu} &=& 
[(A_{1}w+A_{2})\dot{H} \ + \ A_{1}rH']\delta_{\mu}^{w}\delta_{\nu}^{w} \nonumber \\[2mm]
&&+ \ 2A_{1}g_{\mu \nu} \, .
\label{33.kj}
\end{eqnarray}
For $X^{\rm c}$ to be a full-fledged CKVF, the right-hand side must be proportional to the metric tensor (or, by taking trace of the left-hand side of~(\ref{33.kj}), equal to $\frac{1}{2} g_{\mu \nu} \nabla^{\nu} X^{\rm c}_{\nu} $), in which case the condition
\begin{eqnarray}
(A_{1}w+A_{2})\dot{H} \ + \ A_{1}rH' \ = \ 0\, ,
\label{34.vg}
\end{eqnarray}
must hold. Now we can consider $H(w,r)$ as given by~(\ref{H}). Eq.~(\ref{34.vg}) then leads (after multiplying by $r$)  to the third-order equation in $r$ of the form
\begin{eqnarray}
0&=& \ r^{3}[F_{4}'f(w) \ + \ 2A_{1}F_{4}] \ + \ r^{2}[F_{3}'f(w) \ + \ A_{1}F_{3}]\nonumber \\[2mm]
&&+ \ rF_{1}'f(w)  \ + \
F_{2}'f(w) \ - \ A_1F_{2}  \, ,
\end{eqnarray}
where $f(w)=A_{1}w+A_{2}$. Since both $r$ and $w$ coordinates are independent, the coefficient in front of each $r$-monomial must be zero. This gives
\begin{eqnarray}
\label{Ksym}
&&\mbox{\hspace{-12mm}}F_{1}(w) \ = \ a_{1}\, , \;\;\;\;  F_{2}(w) \ = \ a_{2}(A_{1}w+A_{2})\, , \nonumber \\[2mm]   &&\mbox{\hspace{-12mm}}F_{3}(w) \ = \ \frac{a_{3}}{A_{1}w+A_{2}}\, , \;\;\;\; F_{4}(w)\ = \ \frac{a_{4}}{(A_{1}w+A_{2})^{2}}\, ,
\end{eqnarray}
where $a_{i}$ are real constants. The first branch solution (\ref{vacback}) automatically has $F_{1}$ and $F_{2}$ in the form compatible with~(\ref{Ksym}), so $X^{\rm c}$ generates conformal isometry provided $F_{3}$ and $F_{4}$ satisfy~(\ref{Ksym}). On the other hand, the solution (\ref{vacfull}) must satisfy (\ref{Ksym}) for $F_{1}$ and $F_{2}$, and if it does, $F_{3}$ and $F_{4}$ are automatically of the correct form.

In GR, there are several types of Weyl (plus coordinate) transformations that can bring the line element of LMC into a {conformally static} form. Here we present an explicit example known from GR~\cite{Vhorizons}  that works just as well in WCG. To this end, we observe that the coordinate transformation $R^{2}= r/(A_1w+A_2), \: T=\ln( A_1w+A_2 )$ allows us to cast the line element~(\ref{metrika}) in the form 
\begin{eqnarray}
ds^{2} &=& \ e^{2T}\left\{\left[\frac{H(T,R)}{A_1^{2}} \ + \ \frac{2cR^{2}}{A_1} \right]dT^{2} \ + \ \frac{4cR}{A_1}dTdR \right. \nonumber \\[2mm] 
&&+ \left. \ R^{4}d\Sigma^{2}\right\}\, ,
\end{eqnarray}
where $H(T,R) \equiv H(r(T,R), w(T))$ and $d\Sigma^{2}$ stands for the $(x,y)$ sector in~(\ref{metrika}). For LMC, we have 
\begin{eqnarray}
H(R,T) = 
 a_{1} \ + \  \frac{a_{2}}{R^{2}} \ + \ a_{3}R^{2} \ + \ a_{4}R^{4} \ \equiv \  H(R)\, ,~~~
\end{eqnarray}
where $a_i$ are defined in~(\ref{Ksym}).  Consequently, the Weyl transformed metric 
\begin{eqnarray}
\label{Cstatic}
ds^{2}  &=&  \left[\frac{H(R)}{A_1^{2}} \ + \ \frac{2cR^{2}}{A_1} \right]dT^{2} \ + \ \frac{4cR}{A_1} \ \! dTdR \nonumber \\[2mm] 
&&+ \ R^{4}d\Sigma^{2} ,
\end{eqnarray}
is clearly static with KVF~(\ref{KVF}) of the form $X^{\rm c}=\partial_{T}$. The corresponding Killing horizon(s) are therefore located at $g_{TT}=0$ in these coordinates. These locations, generally different from trapping horizons, are important, for example, for thermal radiation or for event horizon localization~\cite{wald2}. In particular, in the original dynamical spacetime it is typically complicated to locate the event horizon, in static spacetime it is comparatively easier. Note that by another coordinate transformation, namely $T=\tilde{w}A_1$, $R^{2}=\tilde{r}$, we can bring~(\ref{Cstatic}) back into the form that is formally identical to~(\ref{metrika}), with the proviso that $H$ depends only on $\tilde{r}$. The latter clearly coincides with M-K spacetime or its topological black hole analogues.



Since the solutions~(\ref{vacback}) and~(\ref{vacfull}) are locally Weyl equivalent to M-K (or WCG topological black hole) solutions, which are locally equivalent to Einstein spaces (see Sec.~\ref{Sec.III.op}), we expect our solutions to be at least locally equivalent to Einstein spaces as well. This means that there should exist a Weyl scaling function $\Omega$ such that $\tilde{g}_{\mu \nu} =\Omega^{2}(r,w)g_{\mu \nu} $, where $g_{\mu \nu} $ is given by (\ref{metrika}) with (\ref{H}) satisfying the GR equation $\tilde{G}_{\mu \nu}+\Lambda \tilde{g}_{\mu \nu}=0$. In this context, we first note that the $rr$-component of the Einstein tensor of such a metric is easy to compute, and it reads 
\begin{eqnarray}
\tilde{G}_{rr} \ = \ \frac{[4 (\Omega')^{2}-2 \Omega \Omega'']}{\Omega^{2}} \ = \ 2 \Omega \left(\frac{\Omega'}{\Omega^2} \right)'\,  .
\end{eqnarray}
By using the fact that $g_{rr} = 0$, the GR equation implies a Riccati equation
\begin{eqnarray}
\Omega' \ = \ -f_1(w) \ \!\Omega^{2},
\end{eqnarray}
which has the general solution
\begin{equation}
\label{omega}
\Omega(r, w) \ =  \frac{1}{f_{1}(w)r \ + \ f_{2}(w)}\, . 
\end{equation}
To find the functions $f_{1}(w)$ and $f_{2}(w)$, we would have to analyze other components of the GR equation. This is quite a technically challenging task, but since the evaluation requires only ordinary differentiation, we can entrust the entire calculation to a symbolic computational method. We performed the actual calculations using Maple 2024. The resulting explicit forms of the components of the Einstein tensor employing~(\ref{omega}) along with the solutions for $f_{1}(w)$ and $f_{2}(w)$ are relegated to the Appendix A. The main results of Appendix A are 
\begin{itemize}
  \item For any $F_1(w)$, $F_2(w)$, and $C_1$ from the branch~(\ref{vacfull}), there always exist functions $f_{1}(w)$ and $f_{2}(w)$ [given by relations (\ref{cEv1a}) or (\ref{cEv1b})].
  \item The existence and number of problematic points depends on the form of $F_i$. In general, they occur at points where $[K+F_1(w)]r+3F_2(w)=0$.
  \item For the branch~(\ref{vacback}), $f_{1}(w)$ and $f_{2}(w)$ can also be found. They can be formulated either in explicit forms [given by the relations (\ref{cEv2a}) or (\ref{58.vb})] or in the form of differential equations, cf. Eq.~(\ref{cEv2b}) [which may generally contain problematic points for $\Omega(r,w)$].
\end{itemize}
\begin{itemize}
  \item All $\Omega(r,w)$ used for transformation to Einstein spaces are either of Riegert type or of (\ref{OmT}) type, i.e. all these spaces admit (at least) one non-null KVF.
  \item There is a large class of VS-type metrics in WCG that have no non-singular Weyl transformations to Einstein spaces, i.e. they are only locally Weyl equivalent to Einstein spaces.
\end{itemize}
%
%
Problematic points are of course of interest, because they precisely refer to situations where the Weyl equivalence between vacuum WCG and Einstein spaces is broken and both theories develop different global causal structures. Weyl inequivalence also often indicates that the two spacetimes have different topologies~\cite{CP-N}. This is because topological invariants, such as homotopy groups or Betti numbers, do not change under non-singular Weyl rescaling, while they might change under singular Weyl transformation. The same reasoning applies to local Weyl equivalence to static solutions. It should be noted, however, that despite these problems, there are still some properties that can be sensibly transferred between locally equivalent spaces [see, e.g., the previously mentioned horizons and their thermodynamics \cite{VT}].

Before we close this section, there is one more conformal transformation worth mentioning. In particular, by introducing a new coordinate $\tilde{r} = 1/r$, we obtain from (\ref{metrika}) and (\ref{H}) a line element of the form 
\begin{eqnarray}
   ds^{2}&=&\tilde{r}^{2}
   \{[F_{1}(w)\tilde{r}^{2}+F_{2}(w)\tilde{r}^{3}+F_{3}(w)\tilde{r}+F_{4}(w)]du^{2} \nonumber \\[2mm]
   &&+ \ 2cdud\tilde{r} \ + \ d\Sigma^{2}\}\, .  
\end{eqnarray}
This shows that our metric is locally Weyl equivalent to a {\em Kundt spacetime}~\cite{uč}, with conformal factor $\Omega=\tilde{r}$. This does not really seem to be too surprising, since conformal Kundt coordinates are used to find and study black hole solutions in the GR as well as in modified gravity theories~\cite{CKundt}. In conventional GR a Kundt geometry is often used as model for gravitational waves and can represent scenarios where energy is radiated away from a source, contributing thus to the study of gravitational radiation~\cite{uč}.

\vspace{3mm}
\section{Non-vacuum solutions \label{Sec.V.bn}}

In this section we extend the Bach vacuum case, studied in Sec.~\ref{Sec.IV.dc} by considering some special $T^{\mu \nu}$, the choice of which will be inspired by the standard VS form~(\ref{V}). In particular, we will focus here on $T^{\mu\nu}$, which describes the Coulomb field  and null dust. Other possibilities such as, e.g., the scalar field~\cite{scalar} will be relegated to our future work.

The general action, including WCG, electromagnetic field, and potentially some other conformal matter field, can be written in the form
\begin{equation}
S \ = \ S^{\text{w}} - \  \frac{1}{4 e^{2}}\int d^{4}x \sqrt{-g}  \ F_{\mu \nu}F^{\mu \nu}  \ + \ S^{M}\, ,
\end{equation}
where $e^{2}$ is the electromagnetic coupling constant, $S^{\text{w}}$ corresponds to (\ref{akce}), and $S^{M} $ is a conformal matter action.

First, let us add the Coulomb field, i.e. $F_{wr}=q(w)/r^{2}$, which gives us for the metric~(\ref{metrika}) and the electromagnetic energy-momentum tensor 
\begin{eqnarray}
T_{\mu \nu}^{EM} \ = \ \frac{1}{e^{2}}\left[F_{\mu \lambda}F_{\nu}^{~\lambda} \ - \ \frac{1}{4} g_{\mu \nu} F^{\alpha \beta}F_{\alpha \beta}\right]\, ,
\end{eqnarray}
the components (only non-zero components are listed)
\begin{eqnarray}
\label{TEM}
&&\mbox{\hspace{-4mm}}T^{EM}_{xx} \ = \ T^{EM}_{yy} \ = \ \frac{8q^{2}(w)}{e^{2} r^{2}(4+K(x^{2}+y^{2}))^{2}}\, , \nonumber \\[2mm]
&&\mbox{\hspace{-4mm}}T^{EM}_{rw} \ = \ -\frac{q^{2}(w)c}{2e^{2}r^{4}}\, , \quad T^{EM}_{ww}\ = \ \frac{-q^{2}(w)H(r,w)}{2e^{2}r^{4}}\, .~~~~~~
\end{eqnarray}
Note in particular that the $rr$-component is zero, so Eq.~(\ref{Brr}) [and thus also form of (\ref{H})] holds unchanged.

When adding pure radiation/null dust, we might assume that the $T^M$ tensor has only the non-zero $ww$-component. The latter automatically guarantees that the ensuing $T^M_{\mu\nu}$ is traceless, in fact 
\begin{eqnarray}
(T^M)_{~\mu}^{\mu} \ = \  g^{ww}\ \!T^M_{ww}  \ = \  0\, .
\end{eqnarray}
We can also assume that $T^{M}_{ww}$ has the form $T^{M}_{ww}=A(w)/r^{2}+B(w)/r^{3}$, because after multiplying by $r^{5}$ --- as in $B_{ww}$ in~(\ref{Bww}) --- this is of the 2nd and 3rd order in $r$. Since Eq.~(\ref{Bwr}) must still hold, we do not want a new function to be of 0th and 1st order in $r$. The corresponding independent equations for Bach's non-vacuum case~(\ref{nvac}) with Coulomb field and null dust thus are
\begin{widetext}
\begin{eqnarray}
\label{nonvac eqs}
&&K^{2} \ + \ 3F_{2}F_{3} \ - \ F_{1}^{2} \ + \ 6cF_{2}' \ = \  \alpha q^{2}\, , \nonumber  \\[2mm]
&&K^{2}F_{3} \ - \ F_{3}F_{1}^{2} \ + \ 3F_{2}F_{3}^{2} \ - \ 2cF_{1}'F_{1} \ + \ 3cF_{3}'F_{2} \ + \ 9cF_{3}F_{2}'\ + \ 6F_{2}''\ = \ \alpha q^{2} F_{3} \ + \ B\, , \nonumber \\[2mm]
&&K^{2}F_{4} \ - \ F_{4}F_{1}^{2} \ + \ 3F_{2}F_{3}F_{4} \ + \ 3cF_{2}F_{4}'\ - \ cF_{3}F_{1}' \ + \ 12cF_{4}F_{2}'\ + \ 2F_{1}'' \ = \ \alpha q^{2} F_{4}\ + \ A\, .
\end{eqnarray}
\end{widetext}
Here we have set $\frac{3}{2}G_{{\text{w}}}^{2}/{e^{2}} = \alpha$. 

We start solving~(\ref{nonvac eqs}) for the $F_{2}(w)=0$ branch of the solutions. To do so, we express $F_{1}(w) $ from the first equation, and insert the solution $F_{1}(w)=\pm \sqrt{K^{2}-\alpha q^{2}(w)}$ into the remaining equations. Since we are going to divide by $q'(w)q(w) $ in the next step, let us first consider the possibility that this term is zero, i.e., that $q(w) $ is constant. In this case, the 2nd equation forces $B(w)=0 $ and the 3rd $A(w)=0 $, while the remaining functions are not touched. So, the first class of solutions is 
\begin{eqnarray}
\label{nonvacback}
&&q(w)\ = \  q\, , \;\; A(w) \ = \ B(w) \ = \ F_{2}(w) \ = \ 0\,  , \nonumber  \\[2mm]
&&F_{1}(w) \ = \ \pm \sqrt{K^{2}- \alpha q^{2}}\, , 
\end{eqnarray}
and $F_{3}(w)$ and  $F_{4}(w)$ are  arbitrary. This is the most general electrovacuum ``background'' solution that is the non-vacuum analog of~(\ref{vacback}). Note that this imposes a condition on the value of $q$, since the expression under the square root must be positive.   

If $q(w)$ is not constant, we can express $F_{3}(w)$ from the 3rd equation of~(\ref{nonvac eqs}) with a complete solution as
\vspace{4mm}
\begin{eqnarray}
\label{nonvacx} 
F_{1}(w) &=& \pm  \ \sqrt{K^{2}-\alpha q^{2}(w)}\, , \;\; B(w) \ = \  2c\alpha q(w)q'(w)\, ,  \nonumber \\[2mm]
F_{2}(w) &=& 0\, , \nonumber \\[2mm]
F_{3}(w) &=& \frac{-4\alpha^{2} q(w)q''(w) +2K^{2}cB'(w) }{F_{1}^{2}(w)B(w)} \nonumber \\[2mm]
&&\pm  \ \frac{ 2A(w)F_{1}^{3}(w)}{F_{1}^{2}(w)B(w)},\\[0.9mm] \nonumber
\end{eqnarray}
where $F_{4}(w)$, $A(w)$ are  arbitrary,  $q(w) $ is arbitrary apart from the conditions $q(w) \neq const.$ and $K^{2}/\alpha \geq q^{2}(w) $  (notice, however that for discrete points $w$, where $K^{2}/\alpha = q^{2}(w) $ solution diverges). This solution has no analog in vacuum solutions.

Finally, we focus on the $F_{2}(w)\neq 0$ branch of solutions.  From the 1st equation of~(\ref{nonvac eqs}) we express $F_{3}(w)$ and insert into the 2nd equation, from which it is easy to express $B(w)$. The 3rd equation represents a first-order linear differential equation for $F_{4}(w)$, which in general can only be integrated by quadratures. The solution is
\begin{widetext}
\begin{eqnarray}
\label{nonvac full} 
F_{3}(w)&=&\frac{-K^{2}+F^{2}_{1}(w)-6cF_{2}'(w)+\alpha q^{2}(w)}{3F_{2}(w)}\, ,\;\; \; B(w) \ = \ 2\alpha cq(w)q'(w)\, , \nonumber \\[2mm]
F_{4}(w)&=&\frac{\int [\alpha q^{2}(w)F_{1}'(w)+3cA(w)F_{2}(w)]dw+C_{2}+\frac{F_{1}^{3}(w)}{3}-6cF'_{1}F_{2}-K^{2}F_{1}}{9F^{2}_{2}(w)}\, ,~~~~~~~~
\end{eqnarray}
\end{widetext}
where $F_{1}(w)$, $F_{2}(w) \neq 0$, $q(w)$ and $A(w)$  are  arbitrary and $C_{2}$ is an integration constant. (Again, the case in which $F_2(w)=0$ at some discrete points belongs to this branch, but diverges at those points.) This solution represents a non-vacuum counterpart of the solution (\ref{vacfull}). It can be integrated directly in the case of electrovacuum, i.e. when $ A(w)=B(w)=0$ (and thus $q(w) = const.$), yielding
\begin{widetext}
\begin{eqnarray}
\label{elecktrovacuum}
 F_{3}(w) &=& \frac{-K^{2} \ + \ F_{1}(w) \  - \ 6cF_{2}'(w) \ + \ \alpha q^{2}}{3F_{2}(w)}\, , \;\; \; A(w)\ = \ B(w) \ = \ 0\, , \;\;\;  q(w) \ = \ q\, , \nonumber \\[2mm]
F_{4}(w)&=&\frac{-\frac{K^{2}F_{1}(w)}{9} \ + \ \frac{F_{1}^{3}(w)}{27} \ + \ \frac{\alpha q^{2}F_{1}(w)}{9} \ - \ \frac{2cF_{2}(w)F_{1}'(w)}{3} \ + \ C_{2}}{F_{2}^{2}(w)}\, ,
\end{eqnarray}
\end{widetext}
where again $F_{1}(w)$  and $F_{2}(w) \neq 0$ are arbitrary and $C_{2} $ is an integration constant.
It is easy to check that Eq.~(\ref{elecktrovacuum}) reduces directly to~(\ref{vacfull}) for $q=0$.

When discussing some of the basic properties of non-vacuum solutions, we can take advantage of the fact that the results in Sec.~\ref{Sec.IV.dc} are mostly computed for the general metric (\ref{metrika}) with $H(r,w)$ given by~(\ref{H}). 
Since non-vacuum solutions have the same metric structure as the vacuum ones (only the explicit form of the functions $F_i(w)$ is affected by the presence of $T^{\mu\nu}$), they are also (at most) of algebraic type D. In fact, the only possibility where solutions are conformally flat (i.e. Petrov type O) is already included in the vacuum case (\ref{vacback}), so all nontrivial non-vacuum solutions are type D. In all these nontrivial cases, there is a singularity at $r=0$ in the invariants~(\ref{I}) and~(\ref{C2}). Possible singularities in $w$ caused by discrete points where the denominators in~(\ref{nonvacx}) or~(\ref{nonvac full}) are zero seem to be only coordinate singularities. The trapping horizon(s) remain at $ H(r,w)=0 $, the $r$-roots of which are still at most 3, but their time dependence is now different. 
This is in accord with the charged version of M-K solution~(\ref{M-K}), where the charge only changes the $1/r$ component of the metric and does not add the $1/r^{2}$ component as in GR. For constant $F_i$ functions, solutions (\ref{nonvacback}) and (\ref{elecktrovacuum}) reduce to the charged version of the M-K (or WCG topological) black hole.

Based on the Weyl transformations discussed in Sec.~\ref{IV.C.cc}, we can consider conformal staticity. Birkhoff--Riegert's theorem~\cite{Riegert} has been stated and proved for the electrovacuum situation, so we expect solutions (\ref{nonvacback}) and (\ref{elecktrovacuum}) to be locally Weyl equivalent to static spacetime [charged version of the M-K (or WCG topological) black hole]. In fact, for (\ref{elecktrovacuum}) there is the same CKVF~(\ref{CKVFg}) as for the vacuum case, which becomes KVF when the metric is divided by $r^{2}$. In the case of (\ref{nonvacback}) the metric divided by $r^{2}$ has $(w,r)$ 2-space of constant scalar curvature $\tilde{R}=2F_1 $, so it is a maximally symmetric 2-space, allowing 3 KVF (at least one non-null). In contrast, solutions (\ref{nonvacx}) and in general (\ref{nonvac full}) do not admit CKVF (nor KVF for the metric divided by $r^{2}$), so in general they are not even locally conformally static.

 As for LMC discussed in Sec.~\ref{Sec.II}, the vector field (\ref{KVF}) is again CKVF for the solution~(\ref{nonvacback}) in the special case $F_{3}(w)=a_{3}/(A_1w+A_2)$, $F_{4}(w)=a_{4}/(A_1w+A_2)^{2}$, see Eq.~(\ref{Ksym}). On the other hand, for the solution~(\ref{nonvacx}), where $q(w)$ is not constant, (\ref{KVF}) cannot be CKVF. The solution~(\ref{nonvac full}) has the conformal flow generated by (\ref{KVF}) provided that $F_{1}(w)$ and $ q(w) $ are constant, $ F_{2}=a_{2}(A_1w+A_2) $ and $ A(w)=0 $, i.e. in the special case of electrovacuum~(\ref{elecktrovacuum}). This situation should be contrasted with the situation in GR, where this symmetry (or even the solution) does not exist without radiation.

Similarly as in Sec.~\ref{IV.C.cc}, we can again ask whether there are solutions of Einstein equations that are Weyl equivalent to our Bach (non-vacuum) solutions. To answer this question, we should find under what conditions exists a Weyl scaling function $\Omega(w,r)$ such that $\tilde{g}_{\mu \nu} =\Omega^{2}(r,w)g_{\mu \nu} $, where $g_{\mu \nu} $ is given by (\ref{metrika}) with (\ref{nonvacback})-(\ref{nonvac full}) satisfying the GR equation $\beta^{2}(\tilde{G}_{\mu \nu}+\Lambda\tilde{g}_{\mu \nu})=\tilde{T}_{\mu \nu}$, where $\tilde{T}_{\mu \nu}=\Omega^{-2}T_{\mu \nu}$ and $\beta^{2}$ is the GR coupling constant. Since $T_{\mu \nu}$ has the $rr-$component equal to zero, the form of $\Omega $ given by (\ref{omega}) must still hold. Using this, we can see that $T^{EM}_{\mu \nu}$ (\ref{TEM}) is automatically excluded by the form of the Einstein tensor (see Appendix A), as can be easily seen by comparing the $r$-orders on both sides of Einstein's equation.
So, without loss of generality, we can only consider the case where $q(w) = 0$.  However, zero $q(w)$ automatically implies that $B(w)=0$ in all cases (\ref{nonvacback})-(\ref{elecktrovacuum}). Thus, only pure radiation $A(w)$ could potentially provide a Weyl equivalent solution. This is difficult to check analytically, since the equations with general $A(w)$ are very complicated, but seem to be solvable with mathematical software [best strategy is to compute with general (\ref{H}), (\ref{omega}) and $A(w)$, solving for all $F_{i}(w), f_{i}(w)$ and $A(w)$, and then compare the results with (\ref{nonvac full})]. The results obtained by Maple 2024 symbolic tensor computations show that there are no solutions for non-trivial $A(w)$. Thus, while all Bach vacuum solutions for general E-F spacetimes~(\ref{metrika}) in WCG are at least locally Weyl equivalent to some Einstein space (i.e., vacuum solution of GR equations with cosmological constant), this is not the case for non-vacuum Bach equations. Consequently, non-vacuum WCG is causally different from ordinary GR with the same $T_{\mu \nu}$. This opens new vistas for novel types of cosmological solutions, which might be particularly relevant for the early Universe cosmology, where WCG is expected to be  most pertinent. 


\section{Discussion and conclusion \label{Sec.VI.cc}}

In this paper, we employed the general E-F-like coordinates known from Einstein's general relativity to explicitly derive and analyze Vaidya-type solutions for dynamical spherical, planar and hyperbolic sources in Weyl conformal gravity. This was done both with and without the presence of the Coulomb field and null dust. While static solutions in WCG have already been studied in the literature, a systematic investigation of non-static solutions has been lacking. Here we aimed to address this gap. In this respect, the E-F coordinates provided a powerful technical framework for conducting such a systematic analysis.

First, we have demonstrated the efficiency of these E-F-like coordinates by focusing on vacuum solutions. In particular, we were able to find and classify all solutions of the Bach vacuum equation of the Vaidya type, which, unlike in GR, have a non-trivial (and generally non-static) structure. They are characterized by (at most) two differentiable real functions and one real constant. In the special case where these functions are chosen to be constant, we obtain the celebrated M-K solution or its topological black hole analog.   We found that all considered solutions correspond to spacetimes of Petrov type-D character (or type-O if conformally flat). They can have both a singularity and trapping horizons, the existence of which depends on the choice of the two defining functions and the constant. In analogy to other solutions known from both GR and WCG, the solutions obtained here can represent vacuum dynamical spacetimes. Some of these solutions are purely cosmological (lacking a ``mass'' term), while others correspond to black (or white hole) types that change their horizon positions in time without radiating anything. The same kind of behavior has been observed in~\cite{Smirnov}.  This could have interesting cosmological implications and provide valuable insights into black hole thermodynamics. 

 A natural question that follows from the results presented here is how elements of our complete set of solutions (both vacuum and non-vacuum) can be mutually related via Weyl transformations, or in other words, what kind of inequivalent WCG configurations we have obtained.  In this context, we discussed the Birkhoff--Riegert theorem, which we extended to both hyperbolic and planar cases, and found Weyl transformations that can give solutions to static M-K or WCG topological black hole forms. However, in most cases these transformations contain singular points, i.e. the Weyl equivalence to static solutions is only local. This typically means that various global properties, such as Noether--Wald charge, homotopy or cohomology groups, are different in such solutions.  (However, even in cases where the equivalence is global, these results can still be of interest because they provide a new framework in which it is possible to see and interpret M-K and WCG topological black hole solutions.) As an example of this procedure we used LMC, i.e. a simple analogue of special GR-VS, which is conformally static. It has been shown that a similar conclusion holds for the Weyl equivalence between WCG and Einstein spaces, although in most cases it is only local. The existence of problematic points in the Weyl transformation is of interest, of course, because they refer precisely to situations where the Weyl equivalence between vacuum WCG and Einstein spaces is often broken and both theories develop different global causal structures.


Second, we also extended our analysis to special Bach non-vacuum cases by considering the energy-momentum tensor of the Coulomb field and null dust. We found and classified the corresponding non-vacuum solutions and briefly discussed their basic properties in close analogy to the vacuum case (most of them apply in the same way). The difference lies in the Weyl transformations used. Only the electrovacuum solutions are (locally) conformally static (in accordance with the Birkhoff--Riegert theorem), the solutions which allow for null dust are not. Also, there is no Weyl transformation that would transform non-vacuum solutions into GR spacetimes. As a consequence, non-vacuum WCGs are causally different from standard GR. 


Let us briefly mention other research directions that could advance the present results. Since we have obtained a whole class of solutions, it is rather difficult to say much about their generic applicability. The class includes spacetimes with or without singularity, with or without various types of horizons, physical and non-physical cases, etc. However, there is the possibility of a deeper analysis of some subclasses or special cases, such as LMC, which, similarly to GR, provides a promising toy model system, allowing to discuss collapse and evaporation, number and types of horizons, details of spacetime structure, thermodynamics, etc.
Another interesting possibility would be to use the Newman--Janis formalism~\cite{Crot, J-N+} or its appropriate generalization to rotate the solutions obtained and find associated axisymmetric solutions. This would provide a rather unusual but very instructive way of generating in WCG analogues of the Kerr--Vaidya and Kerr--Newman--Vaidya solutions~\cite{KV}.  
Also, it is possible to extend~(\ref{metrika}) by promoting constants to coordinate functions. Although this seems to be (especially in the case of functions of $r$) complicated, we already obtained results in wormhole-like metrics for some very simple cases which we aim to study further.
For more advanced solutions and their classification, the Newman--Penrose formalism~\cite{CP-N} could ameliorate technical difficulties. Last but not least, another choice of conformally compatible energy-momentum tensors might be interesting, as they would provide yet another WCG analogy to generalized VS. In this respect, the conformally coupled scalar field might serve as an one of important testbed examples.


\section*{Acknowledgments}
We thank Philip Mannheim for enlightening discussions.
P.J. was supported by the Czech Science Foundation Grant (GA\v{C}R), Grant No. 25-18105S.
T.L acknowledges the support from the Grant Agency of the Czech Technical University in Prague, grant No. SGS22/178/OHK4/3T/14.


\appendix

\section{Einstein spacetimes}

Non-zero components of the Einstein tensor for the general metric
(\ref{metrika}) are

\begin{eqnarray}
\label{Einstein}
&&r^{2}G_{ww} \ = \ -KH-HH'r \ + \ c\dot{H}r \ - \ H^{2}\, , \nonumber \\[2mm] 
&&cr^{2}G_{rw} \ = \ -K \ - \ H'r \ - \ H\, , \nonumber \\[2mm]
&&[4 +  K(x^{2} +  y^{2})]^{2}G_{xx/yy} \ = \ -8r(H''r  + 2H')\, .~~~~
\end{eqnarray}

Using Maple 2024 and the ensuing symbolic tensor computation, we find that the components of the Einstein tensor for general (\ref{H}) and (\ref{omega}) are
\begin{widetext}
\begin{eqnarray}
&&\tilde{G}_{xx} \ = \ \tilde{G}_{yy} \ = \ \frac{16r}{[4+K(x^{2}+y^{2})](f_{1}r+f_{2})^{2}}  \ \!\left[ r\left(-f_{1}^{2}F_{1} \ + \ 2f_{1}f_{2}F_{3} \ - \ 3f_{2}^{2}F_{4} \ - \ 6cf_{2}f_{1}' \ + \ 4cf_{2}'f_{1}\right)\right.\nonumber \\[2mm]
&&\mbox{\hspace{9mm}}+\ \left. f_{1}f_{2}F_{1} \ - \ 3f_{1}^{2}F_{2} \ - \ f_{2}^{2}F_{3} \ - \ 2cf_{2}f_{2}' \right]\, , \nonumber \\[2mm]
&&\tilde{G}_{rw} \ = \ \frac{c}{(f_{1}r+f_{2})^{2}r^{2}} \ \! \left[ 2r^{2} \left(2cf_{1}f_{2}' \ - \ 6cf_{2}f_{1}'\ - \ 3F_{4}f_{2}^{2} \ + \ F_{3}f_{1}f_{2}\ - \ Kf_{1}^{2}\right)\right. \nonumber \\[2mm]
&&\mbox{\hspace{9mm}}+ \left.  2r\left(-2cf_{2}f_{2}'-F_{3}f_{2}^{2}-Kf_{1}f_{2}+f_{1}f_{2}F_{1}\right) \ - \  Kf_{2}^{2} \ - \ F_{1}f_{2}^{2} \ + \ 3F_{2}f_{1}f_{2} \right] \, ,\nonumber \\[2mm]
&&\tilde{G}_{ww} \ = \ -\frac{1}{(f_{1}r+f_{2})^{2}r^{3}} \ \! \left[ 120r^{5}\left(KF_{4}f_{1}^{2}\ - \ F_{3}F_{4}f_{1}f_{2} \ + \ cF_{3}f_{1}f_{1}' \ - \ 4cF_{4}f_{1}f_{2}'\ + \ 6cF_{4}f_{2}f_{1}'\ - \ cF_{4}'f_{1}f_{2}\ + \ 3F_{4}^{2}f_{2}^{2} 
\right. \right. \nonumber \\[2mm]
&&\mbox{\hspace{9mm}}- \left. 
 2f_{1}f_{1}''\right) \ + \ 24r^{4}\left(-\ 2F_{1}F_{4}f_{1}f_{2} \ + \ 2cF_{1}f_{1}f_{1}' \ - \ 3cF_{3}f_{1}f_{2}'\ + \ 7cF_{3}f_{2}f_{1}'\ + \ 2cF_{4}f_{2}f_{2}'\ - \ cF_{3}'f_{1}f_{2}\right. \nonumber \\[2mm] 
&&\mbox{\hspace{9mm}}+ \left. \ KF_{3}f^{2}_{1}\ + \ 2KF_{4}f_{1}f_{2} \  - \ 2f_{1}''f_{2}\ - \ 2f_{2}''f_{1} \ - \ F_{3}^{2}f_{1}f_{2} \ + \ 5F_{3}F_{4}f_{2}^{2} \ - \ f_{2}^{2} \ - \ cF_{4}'f_{2}^{2}\right) \nonumber \\[2mm] 
&&\mbox{\hspace{9mm}}+ \ 6r^{3}\left(-3F_{1}F_{3}f_{1}f_{2}\ - \ 2cF_{1}f_{1}f_{2}' \ + \ 8cF_{1}f_{2}f_{1}'\ - \ 3F_{2}F_{4}f_{1}f_{2}\ + \ 3cF_{2}f_{1}f_{1}'\ + \ 3cF_{3}f_{2}f_{2}' \ - \ cF_{1}'f_{1}f_{2}\right.\nonumber \\[2mm]
&&\mbox{\hspace{9mm}} +
 \left. \ KF_{1}f_{1}^{2} \ + \ KF_{4}f_{2}^{2} \ + \ 2F_{3}^{2}f_{2}^{2} \ + \ 2KF_{3}f_{1}f_{2}\ - \ 2f_{2}''f_{2} \ + \ 4F_{1}F_{4}f_{2}^{2}\ - \ cF_{3}'f_{2}^{2}\right) \nonumber \\[2mm] 
&&\mbox{\hspace{9mm}} + \ 2r^{2}\left(KF_{2}f_{1}^{2} \ + \ KF_{3}f_{2}^{2} \ + \ 4cF_{1}f_{2}f_{2}' \ - \  \ 4F_{2}F_{3}f_{1}f_{2} \ - \ cF_{2}f_{1}f_{2}'\ + \ 9cF_{2}f_{2}f_{1}'\ - \ cF_{2}'f_{1}f_{2}\ + \ 2KF_{1}f_{1}f_{2} \right.\nonumber \\[2mm] 
&&\mbox{\hspace{9mm}}- \left. \ 2F_{1}^{2}f_{1}f_{2} \ + \ 3F_{1}F_{3}f_{2}^{2} \ + \ 3F_{2}F_{4}f_{2}^{2} \ - \ cF_{1}'f_{2}^{2}\right) 
\ + \  r\left(KF_{1}f_{2}^{2} \ - \ 5F_{1}F_{2}f_{2}f_{1} \ + \ 5cF_{2}f_{2}f_{2}'\ + \ F_{1}^{2}f_{2}^{2} \right.\nonumber \\[2mm] 
&&\mbox{\hspace{9mm}}+ \left.\left.\ 2KF_{2}f_{1}f_{2} \ + \ 2F_{3}F_{2}f_{2}^{2} \ - \ cF_{2}'f_{2}^{2}\right)\ + \ F_{1}F_{2}f_{2}^{2}\ - \ 3F_{2}^{2}f_{1}f_{2} \ + \ KF_{2}f_{2}^{2}\right]\, .
\label{EinstH}
\end{eqnarray}
\end{widetext}
After inserting (\ref{vacfull}) into equation $ \tilde{G}_{\mu \nu}+ \Lambda \tilde{g}_{\mu \nu} $, we find that 
\begin{eqnarray}
\label{cEv1a} 
f_{1}(w)&=&\frac{(K+F_{1}(w))\sqrt{\Lambda (2K^{3}+27C_{1})}}{2K^{3}+27C_{1}}\, , \nonumber \\[2mm] 
f_{2}(w)&=&\frac{3F_{2}(w)\sqrt{\Lambda (2K^{3}+27C_{1})}}{2K^{3}+27C_{1}}\, ,
\end{eqnarray}
for the case and $2K^{3}+27C_{1} \neq 0$, while for $2K^{3}+27C_{1} = 0$, we have
\begin{eqnarray}
\label{cEv1b}
&&f_{1}(w) \ = \ \lambda_{1}(K+F_{1}(w))\, , \nonumber \\[2mm]
&&f_{2}(w) \ = \ 3\lambda_{1} F_{2}(w)\, , \nonumber \\[2mm]
&&\Lambda \ = \ 0\, ,
\end{eqnarray}
where $\lambda_{1} $ is a constant. In passing, we note that since $f_{1}(w)$ and $f_{2}(w)$ must be real functions, the signature of $\Lambda$ (which is actually part of the solution) must match the signature of the term it multiplies inside the square root. The same comment applies also in Eq.~(\ref{cEv2a}).

For the branch (\ref{vacback}), we find that
\begin{equation}
\label{cEv2a}
f_{1}(w) \ = \ \pm \sqrt{\Lambda K}/K\, , \;\;\;\; f_{2}(w) \ = \ 0\, .
\end{equation}
for $F_{1}(w)=K, K \neq 0$. The case $F_{1}(w)=-K, K \neq 0 $ is more complicated, here it is only possible to find differential relations for $f_{1}(w)$, and $f_{2}(w)$, which can be solved after $F_{3}(w), \: F_{4}(w) $ and $K$ are chosen. These relations read
\begin{widetext}
\begin{eqnarray}
\label{cEv2b}
&&F_{3}(w) \ = \ \frac{ -2Kf_{1}(w) \ - \ 2cf_{2}'(w)}{f_{2}(w)}\, , \quad
F_{4}(w) \ = \ \frac{-3K f_{1}^{2}(w) \ - \ 6cf_{1}'(w)f_{2}(w) \ + \ \Lambda}{3f_{2}^{2}(w)}\,.
\end{eqnarray}
\end{widetext}
Although the equations look quite complicated, one can solve them explicitly for the special case $\Lambda = 0$ by changing to new functions $z(w) = f_1(w)/f_2(w)$ and $y(w) = \log(f_2(w))$.  In such a case, the system~(\ref{cEv2b}) is simplified so that the solution for $z(w)$ and $y(w)$ can be easily written in terms of integrals over $F_3(w)$ and $F_4(w)$.

Finally for case $F_{1}(w)=K=0$ we get
\begin{equation}
    f_{1}(w) \ = \ \lambda_{2}\, ,  \quad f_{2}(w)\ = \ 0\, , \quad \Lambda\ = \ 0\, ,
    \label{58.vb}
\end{equation}

where $\lambda_{2}$ is a constant. 

\section{Graphical summary~\label{Appendix B}}

For the reader's convenience, this appendix summarizes solutions obtained in generalized E-F coordinates and their key properties in Tab.~\ref{tab:solutions}.  When the Weyl equivalence with Einstein and/or static spacetimes is described as ``local'' or ``local'', it applies to the general case. There may be special cases where the equivalence is global (see section ~\ref{IV.C.cc}). Singularities are listed under the assumption that all functions $F_i(w)$ defining certain solutions are bounded and non-singular. Several important relations between solutions obtained in this paper and known WCG solutions --- MK solution and its topological black hole analogies are represented by the web chart in Fig.~\ref{fig:relations}.
\begin{widetext}

\begin{table}[h]
    \centering
    \small
    \renewcommand{\arraystretch}{1.3}
    \begin{tabular}{|c||c|c|c|c|c|c|c|}
    \hline
    \multicolumn{1}{|c||}{
        \begin{tikzpicture}[baseline=(current bounding box.center)]
        \put (-51,16) {\line(3,-0.65){131}}
            \useasboundingbox (0,0) rectangle (1,0.5); 
        \end{tikzpicture}
    } 
    & \multicolumn{3}{c|}{vacuum solutions} & \multicolumn{4}{c|}{non-vacuum solutions} \\
    \cline{2-8}
    & (\ref{vacback}), $F_1 = -K$ & (\ref{vacback}), $F_1 = +K$ & (\ref{vacfull}) & (\ref{nonvacback}) & (\ref{nonvacx}) & (\ref{nonvac full}) & (\ref{elecktrovacuum}) \\
    \hline
    \hline
        Petrov type & O & D & D & D & D & D & D \\
    \hline
        Singularities in $r$ (in invariants) & no & yes & yes & yes & yes & yes & yes \\
    \hline
        Singularities in $w$ (coordinate)  & no & no & possible & no & no & possible & possible \\
    \hline
        Einstein equivalence & local & local & local & none & none & none & none \\
    \hline
        Conformally static & locally & locally & locally & locally & no & no & locally \\
    \hline
    \end{tabular}\\[3mm]
    \caption{Solutions of WCG obtained in generalized E-F metric~(\ref{metrika}).}
    \label{tab:solutions}
\end{table}

\small{

\begin{center}
    \begin{tikzpicture}
\node[draw, thick, minimum width=2cm, minimum height=1cm] (A) at (3,0) {Eq.~(\ref{nonvacback})};
\node[draw, thick, minimum width=2cm, minimum height=1cm] (B) at (6,0) {Eq.~(\ref{nonvacx})};
\node[draw, thick, minimum width=2cm, minimum height=1cm] (C) at (9,0) {Eq.~(\ref{nonvac full})};
\node[draw=none, thick, minimum width=2cm, minimum height=1cm, align=center] (D) at (6,-3) {NO VACUUM \\ LIMIT};
\node[draw, thick, minimum width=2cm, minimum height=1cm] (E) at (9,-3) {Eq.~(\ref{elecktrovacuum})};
\node[draw, thick, minimum width=2cm, minimum height=1cm] (F) at (3,-6) {Eq.~(\ref{vacback})};
\node[draw, thick, minimum width=2cm, minimum height=1cm] (G) at (9,-6) {Eq.~(\ref{vacfull})};
\node[draw, thick, minimum width=2cm, minimum height=1cm, text width=3.2cm, align=center] (H) at (0,-9) {charged MK/WCG topological BH without $1/r$ term};
\node[draw, thick, minimum width=2cm, minimum height=1cm, text width=3.2cm, align=center] 
    (I) at (5,-9) {MK/WCG topological BH without $1/r$ term};\textbf{}
\node[draw, thick, minimum width=2cm, minimum height=1cm, text width=3.2cm, align=center] (J) at (9,-9) {MK/WCG topological BH with $1/r$ term};
\node[draw, thick, minimum width=2cm, minimum height=1cm, text width=3.2cm, align=center] (K) at (14,-9) {charged MK/WCG topological BH with $1/r$ term};
\draw[thick, -{Latex[scale=1.2]}] (A.west) -| (H.north) node[near end, right, font=\small] {$F_i = \textrm{const.}$};
\draw[thick, -{Latex[scale=1.2]}] (A.south) -| (F.north) node[near end, right, font=\small] {$q = 0$};
\draw[thick, -{Latex[scale=1.2]}] (C.south) -| (E.north) 
    node[near end, right, font=\small, align=center] {$q' =0$ \\ $A = 0$};
\draw[thick, -{Latex[scale=1.2]}] (B.south) -| (D.north) node[near end, right, font=\small] {};;
\draw[thick, -{Latex[scale=1.2]}] (E.south) -| (G.north) node[near end, right, font=\small] {$q = 0$};
\draw[thick, -{Latex[scale=1.2]}] (G.south) -| (J.north) node[near end, right, font=\small] {$F_i = \textrm{const.}$};
\draw[thick, -{Latex[scale=1.2]}] (E.east) -| (K.north) node[near end, left, font=\small] {$F_i = \textrm{const.}$};
\draw[thick, -{Latex[scale=1.2]}] (K.west) |- (J.east) node[near end, above, font=\small] {$q = 0$};
\draw[thick, -{Latex[scale=1.2]}] (F.east) -| (I.north) node[near end, left, font=\small] {$F_i = \textrm{const.}$};
\draw[thick, -{Latex[scale=1.2]}] (H.east) |- (I.west) node[near end, above, font=\small] {$q = 0$};
\node[above=1mm] at (A.north) {\textit{(electrovacuum)}};
\node[above=4mm] at (B.north) {\textit{(Coloumb +}};
\node[above=1mm] at (B.north) {\textit{null dust)}};
\node[above=4mm] at (C.north) {\textit{(Coloumb +}};
\node[above=1mm] at (C.north) {\textit{null dust)}};
\node[above=6mm] at (F.west) {\textit{(vacuum)}};
\node[above=6mm] at (G.west) {\textit{(vacuum)}};
\node[above=1mm, xshift=13mm] at (E.east) {\textit{(electrovacuum)}};
\node[below=1mm] at (H.south) {\textit{(electrovacuum)}};
\node[below=1mm] at (I.south) {\textit{(vacuum)}};
\node[below=1mm] at (K.south) {\textit{(electrovacuum)}};
\node[below=1mm] at (J.south) {\textit{(vacuum)}}; 
    \end{tikzpicture}\\[5mm]
    
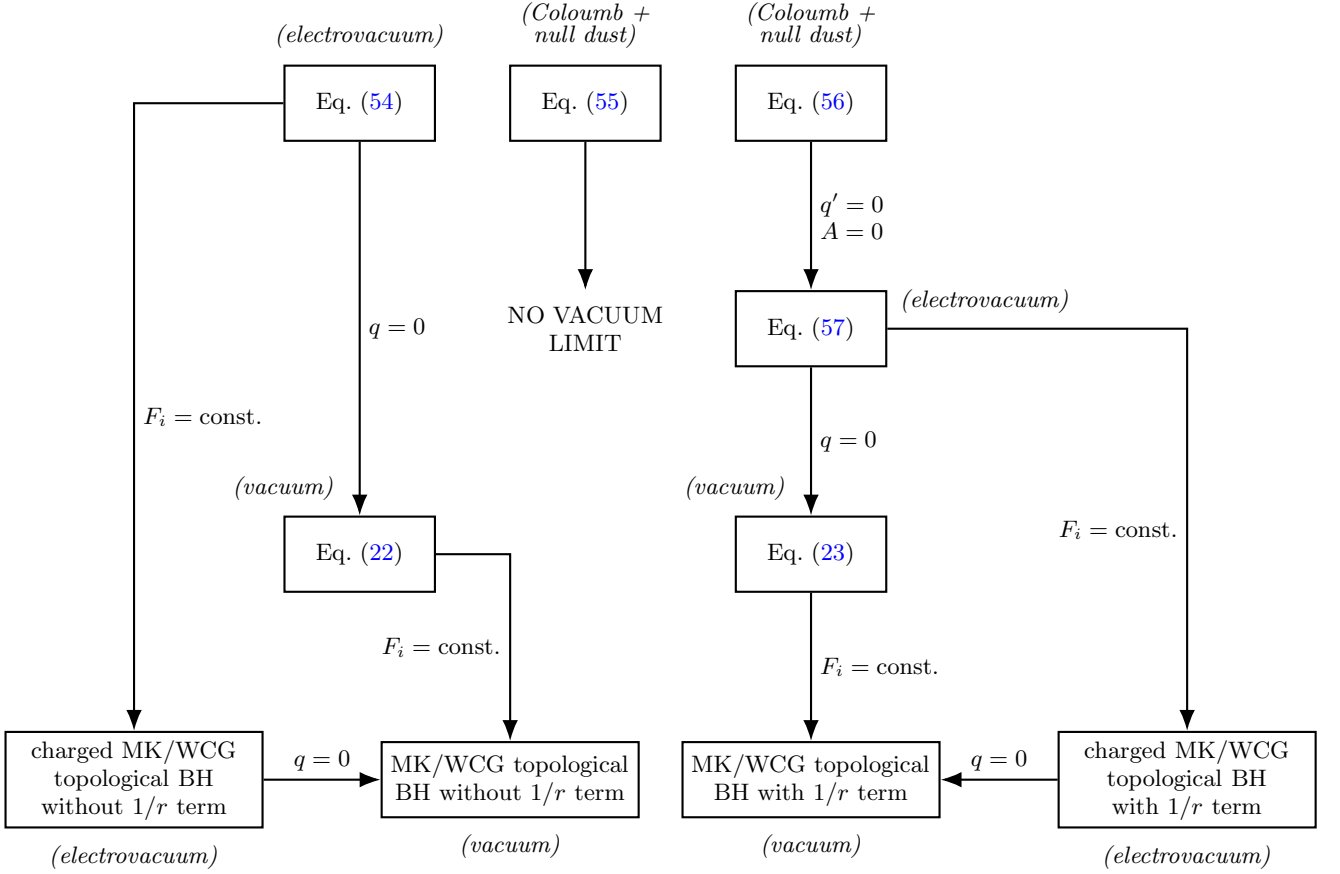
\captionof{figure}{Relations between solutions obtained and known WCG solutions listed in Refs.~\cite{CR-N} and~\cite{Ctop}.}
    \label{fig:relations}
\end{center}

}

\end{widetext}

\vspace{5cm}
\bibliography{basename of .bib file}

\end{document}